\documentclass[12pt]{iopart}

\usepackage{iopams}  
\usepackage{times}

\usepackage{color,colordvi,graphicx}

\begin{document}

\title[Low temperature resonances in D$_2$\,+\,H ]
      {Resonances at very low temperature for the reaction D$_2$\,+\,H}

\author{I. Simbotin and R. C\^ot\'e}

\address{Department of Physics, University of Connecticut, 2152
  Hillside Rd., Storrs, CT 06269, USA} \ead{rcote@phys.uconn.edu}
\vspace{10pt}
\begin{indented}
\item[] December 2016
\end{indented}

\begin{abstract}
  We present numerical results for rate coefficients of reaction and
  vibrational quenching in the collision of H with D$_2(v,j)$ at cold
  and ultracold temperatures.  We explore both ortho-D$_2(j\!=\!0)$
  and para-D$_2(j\!=\!1)$ for several initial vibrational states
  $(v\leq 5)$, and find resonant structures in the energy range
  0.01--10~K, which are sensitive to the initial rovibrational state
  $(v,j)$.  We compare the reaction rates for D$_2$\,+\,H with our
  previously obtained results for the isotopologue reaction
  H$_2$\,+\,D, and discuss the implications of our detailed study of
  this benchmark system for ultracold chemistry.
\end{abstract}

%
%
%
%
%

\section{Introduction}

Research on cold molecules has been growing steadily since they were
first predicted \cite{cote-1997,jmp-1999} and observed experimentally
\cite{pillet-1998,knize-1998}.  Indeed, there is now a vast literature
on cold molecules, including numerous reviews of certain specialized
topics \cite{stwalley:canjchem:04,weck+bala:irpc:06,jeremy:irpc:06,
  jeremy:irpc:07,krems:pccp:08} and several books
\cite{gospel:09,ian.smith:08,w+z:09} dedicated to the physics and
chemistry of cold molecules.  A variety of methods have been developed
to produce cold molecules, thus opening the door for studying cold and
ultracold chemistry, now a rapidly expanding field.  In this article
we explore in detail the reaction D$_2$\,+\,H for energies below
100~K, including the ultracold regime ($T\sim 10^{-6}$~K) which
corresponds approximately to energies from $10^{-10}$ to
$10^{-2}$~eV\@.  We published recently a similar study
\cite{our:NJP-2015:H2D} for the isotopologue reaction H$_2$\,+\,D, and
in this paper we compare and contrast the results for the two
reactions.  In both cases we find resonant features in the energy
range 0.01--10~K, which we identify as shape resonances in the
entrance channel.  We remark that these resonances are very close to
the entrance channel threshold, and thus even small shifts in their
positions can have strong effects on their lineshapes.  Indeed, our
detailed comparison shows that the positions of the resonances for
D$_2$\,+\,H are shifted to slightly higher energy than the resonances
for H$_2$\,+\,D, which is primarily due to the different value of the
reduced mass in the entrance arrangement.  Nevertheless, the
lineshapes of the resonances found for these two reactions are very
different.  Moreover, we also find that the results are strongly
dependent on the initial rovibrational state of the dimer.

The reaction we study in this work, together with its isotopologues,
is of fundamental importance in quantum chemistry, and is also
relevant to astrophysics, especially for the astrochemistry in the
early universe \cite{astro} and for the evolution of cold molecular
clouds in the earliest stages of star formation \cite{astro-cloud}.
Recent experiments with slow collisions between metastable He and
H$_2$ \cite{Narevicius-2012-Science,Narevicius-2014-NatChem,Narevicius-2015-NatChem,Narevicius-2016-NatPhys}
  have detected similar resonances, and studies of this benchmark
  reaction should lead to a better understanding of the energy surface
  and of the relevant scattering processes.
%
In section~\ref{sec:theory} we give a brief description of
the theoretical and numerical tools used, as well as the properties of
this benchmark system. We present and analyze the results in
section~\ref{sec:results}, and conclude in section~\ref{sec:conclusion} .

\section{Theoretical and computational details}
\label{sec:theory}

We consider the benchmark reaction D$_2$\,+\,H at low temperatures and
pay special attention to the effect of the nuclear spin symmetry,
since D$_2$ is a homonuclear molecule.  As is well known, the nuclei
of D$_2$ are spin $i=1$ bosons, with possible values of total nuclear
spin $I$ of 0, 1, and 2, and the symmetrization requirements for the
nuclear wave function of D$_2$ restricts its rotational states.
Specifically, for total nuclear spin $I=2$ (maximal value) and also
for $I=0$, the nuclear spinor part of the wave function is symmetric,
and thus only rotational states $j=0,2,4$, etc., are allowed, which
one customarily refers to as \emph{ortho}-D$_2$, while for $I=1$, the
spinor is antisymmetric and the corresponding rotational states are
$j=1,3,5$, etc., referred to as \emph{para}-D$_2$.  The coupling of
the nuclear spins to the other degrees of freedom is negligible, and
thus ortho- and para-D$_2$ are treated separately. As in our previous
study of H$_2$+D, hyperfine interactions are neglected
\cite{our:NJP-2015:H2D}.

The expression for the state-to-state cross sections, integrated over
all scattering directions, averaged over the initial rotational states
of the reactant dimer, and summed over the final rotational states of
the product, reads
\begin{equation}
\label{eq:sigma}
\sigma_{n'\leftarrow n}(E) = \frac{\pi}{k^{2}_n}
	\sum_{J=0}^{\infty} \left(\frac{2J+1}{2j+1}\right)
	\sum_{\ell = |J-j|}^{J+j}
    \;\;\sum_{\ell'= |J-j'|}^{J+j'}
     \left|\delta_{n'n}\delta_{\ell'\ell} - S^{J}_{n'\ell', \,n\ell}(E)\right|^{2}.
\end{equation}
The generic notation $n=(a v j)$ stands for the arrangement label
``$a$" and quantum numbers $(vj)$ of the rovibrational states of
D$_2$, and $k_n=\sqrt{2\mu_a E_n^{\rm kin}}$ is the initial momentum
($\hbar=1$, atomic units are used), with
$\mu_a=(m_{D_2}^{-1}+m_H^{-1})^{-1}$ the reduced mass of the binary
system D$_2-$H in the initial arrangement $a=1$, $E_n^{\rm
  kin}=E-\varepsilon_n$ the initial kinetic energy for relative
motion, and $\varepsilon_n$ the rovibrational energies.  $E$ is the
(total) collision energy, $J$ is the total angular momentum, and
$\ell$ is orbital angular momentum for the relative motion.  The
primed symbols indicate the corresponding quantities in the exit
channel, with $n'=(a' v' j')$ and orbital angular momentum $\ell'$.
Here, $a'=a=1$ implies that the system remains in the same arrangement
D$_2$\,+\,H, while $a'=2$ means that it evolved into the other
possible arrangement HD\,+\,D, with reduced mass
$\mu_{a'=2}=(m_{HD}^{-1}+m_D^{-1})^{-1}$.  The conservation of the
total angular momentum $J$ ensures that the S-matrix is block diagonal
with respect to $J$, and thus the matrices $S^J$ are obtained
separately for each $J$.

The state-to-state energy dependent rate constants are simply obtained
by multiplying the corresponding cross sections with the relative
velocity $v_{\rm rel}$ in the initial channel,
\begin{equation}
\label{eq:rate}
{\sf K}_{n'\leftarrow n}(E) = v_{\rm rel} \sigma_{n'\leftarrow n}(E) \; ,
\end{equation}
where $v_{\rm rel}= \sqrt{2E_n^{\rm kin}/\mu_a}$. The total rate
constants for quenching $(a'=a)$ and reaction $(a'\neq a)$ are obtained by
summing over the appropriate final states $n'$,
\begin{eqnarray}
 {\sf K}_{avj}^{\rm Q}(E) & = & \sum_{(v'\!,\,j')\neq(v,\,j)} {\sf K}_{av'j'\leftarrow avj}(E) \;, \label{eq:K-Q} \\
 {\sf K}_{avj}^{\rm R}(E) & = & \sum_{a'\neq a,\,v'\!,\,j'} {\sf K}_{a'v'j'\leftarrow avj}(E) \label{eq:K-R} \;.
\end{eqnarray}
Note that $(a'v'j')=(avj)$ corresponds to elastic scattering, which is
discussed in \S~\ref{sec:elastic}.  The numerical results were
obtained using the ABC reactive scattering code developed by
Manolopoulos and coworkers \cite{abc:cpc:2k}.  To ensure fully
converged results, a sufficiently large number of closed channels were
included, and the propagation of the coupled radial equations was
monitored for adequate accuracy.  Technical details of our
computational approach were published elsewhere
\cite{PCCP-H2+D,our:NJP-2015:H2D}.

\section{Results and discussion}
\label{sec:results}

\begin{figure}[b]
\centerline{\includegraphics[width=0.7\textwidth]{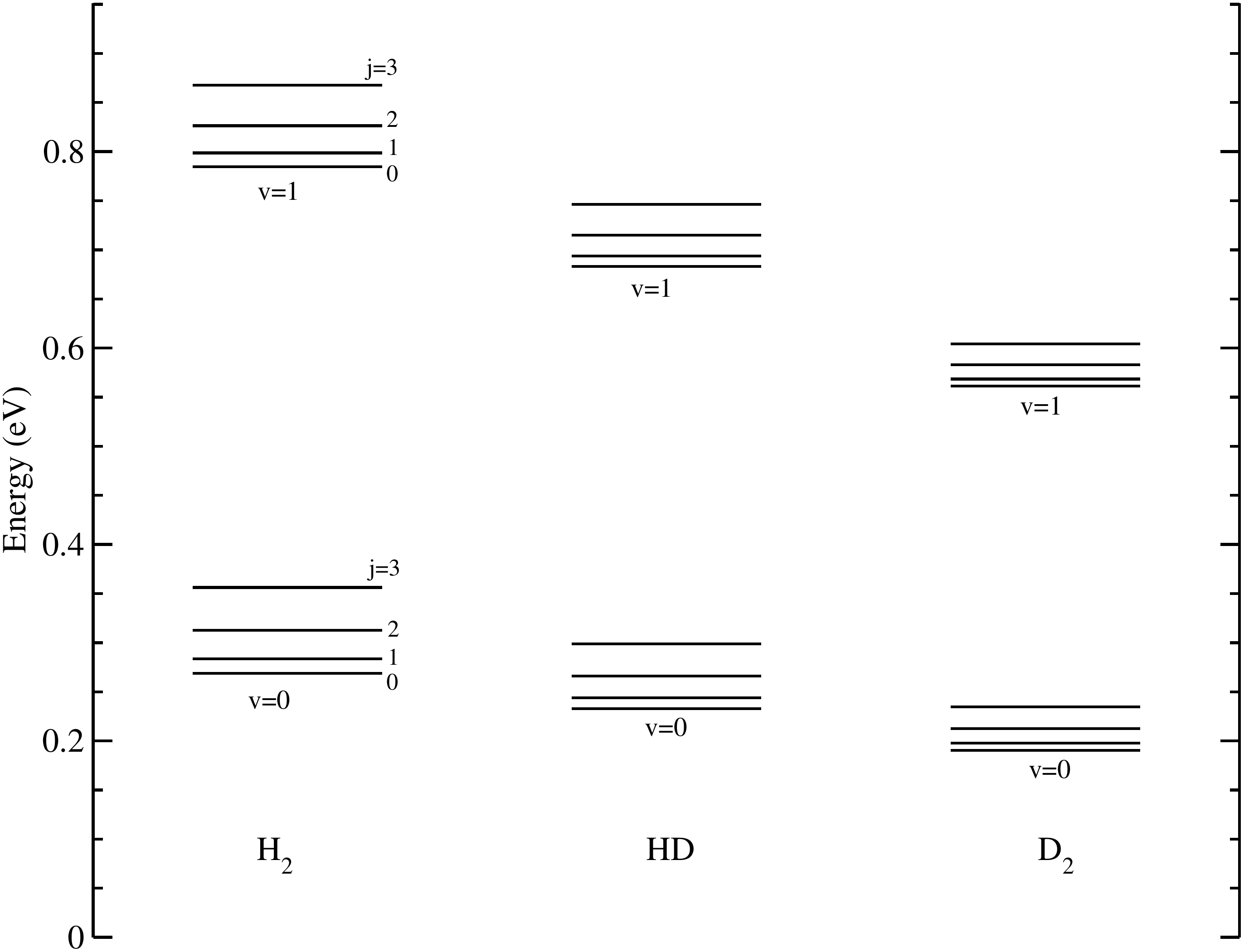}
}
\caption{
Rovibrational energy levels for $v=0,1$ and $j\leq 3$.
\label{fig:rovib-levels}
}
\end{figure}

The benchmark system H$_2$\,+\,D has already been studied in the
ultracold regime \cite{PCCP-H2+D} and also at higher temperatures
\cite{mielke}, while the isotopologue reaction D$_2$\,+\,H has only
recently been explored in the ultracold regime \cite{bala-2016}.
Accurate {\it ab initio} potential energy surfaces exist
\cite{bkmp2:jcp96,mielke} for H$_3$, and we adopt here the electronic
ground state surface of Ref.~\cite{bkmp2:jcp96}, as we did in our
previous studies
\cite{PCCP-H2+D,PRL-rydberg-dressing,our:NJP-2015:H2D}.
Figure~\ref{fig:rovib-levels} shows the rovibrational energy levels
for $v=0,1$ and $j\leq3$, for the reactants H$_2$ and D$_2$, and also
for the product HD\@. Note that the reaction
D$_2(v,j)$\,+\,H\,$\rightarrow$\,HD\,+\,D cannot take place at
vanishing energies for $v=0$ and $j\leq2$, while
H$_2(0,j)$\,+\,D\,$\rightarrow$\,HD\,+\,H does take place.

\subsection{Reaction and quenching rates for j\,=\,0}
\label{sec:j=0}

\begin{figure}[t]
\mbox{}\hfill\includegraphics[width=0.84\textwidth]{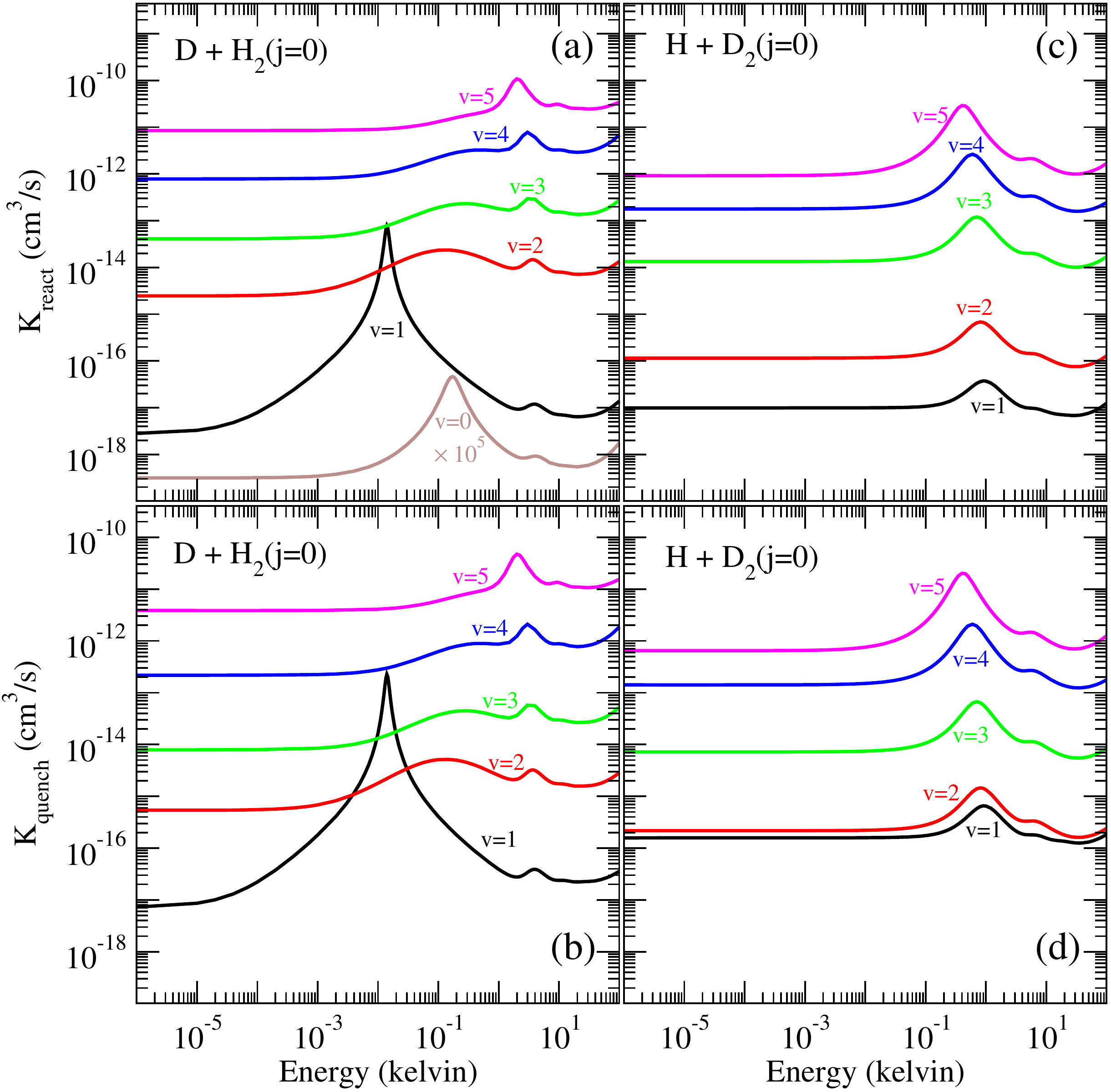}
\caption{
Energy dependent rate coefficients ${\sf K}_{vj}(E)$ for
  $j=0$ and $v\leq 5$.  Upper panels for reaction, lower panels for
  quenching.  The two panels on the left side correspond to H$_2$\,+\,D,
  while the panels on the right to D$_2$\,+\,H, as indicated.
\label{fig:compare:j=0}
}
\end{figure}

In figure~\ref{fig:compare:j=0} we compare the results we obtained
previously \cite{our:NJP-2015:H2D} for H$_2(v,\,j$\,=\,$0)$\,+\,D with
the newly computed results for D$_2(v,\,j$\,=\,$0)$\,+\,H\@.  We
remark that the initial channels $n=(a$\,=\,$1,v,\,j$\,=\,$0)$
correspond to para-H$_2$ and ortho-D$_2$, respectively.  Rate
coefficients for initial vibrational levels $v$\,$\leq$\,5 are
presented; note that $v$\,=\,0 appears only in
figure~\ref{fig:compare:j=0}(a) for
D\,+\,H$_2(0,0)$~$\rightarrow$~HD\,+\,H, which is an exoergic reaction
and will thus occur even at vanishing energies, see
  figure~\ref{fig:rovib-levels}.  ${\sf K}^{\rm Q}_{v=0}$ is absent
in figure~\ref{fig:compare:j=0}(b,~d) because the ground rovibrational
level $(v$\,=\,$0,j$\,=\,$0)$ cannot be quenched any further, while
${\sf K}^{\rm R}_{v=0}$ is absent in figure~\ref{fig:compare:j=0}(c)
because H\,+\,D$_2(0,0)$~$\rightarrow$~HD\,+\,D is endoergic, and all
reaction channels are closed at the low energies considered
here, see figure~\ref{fig:rovib-levels}.

Significant resonant features are readily apparent in
figure~\ref{fig:compare:j=0} in the energy range 0.01--10~K, for all
initial vibrational states for both isotopologue systems.  The most
striking case is the resonant peak for
D\,+\,H$_2(v$\,=\,$1,j$\,=\,$0)$, which stems from a shape resonance
at $E\approx 15$~mK, just above the zero energy threshold. In the case
of D$_2$\,+\,H, this $p$-wave resonance has shifted to $E\approx 1$~K
and is much less pronounced.  Note that for each initial vibrational
channel, the quenching and reaction rates are resonantly enhanced in
identical fashion, which is due to the fact that all resonances
displayed in figure~\ref{fig:compare:j=0} reside in the entrance
channel, near the zero energy threshold, and they will thus imprint
nearly identically on all final channels
\cite{our:jost-NTR:ChemPhys-2015,NTR}.

\begin{figure}[t]
\includegraphics[width=0.47\textwidth]{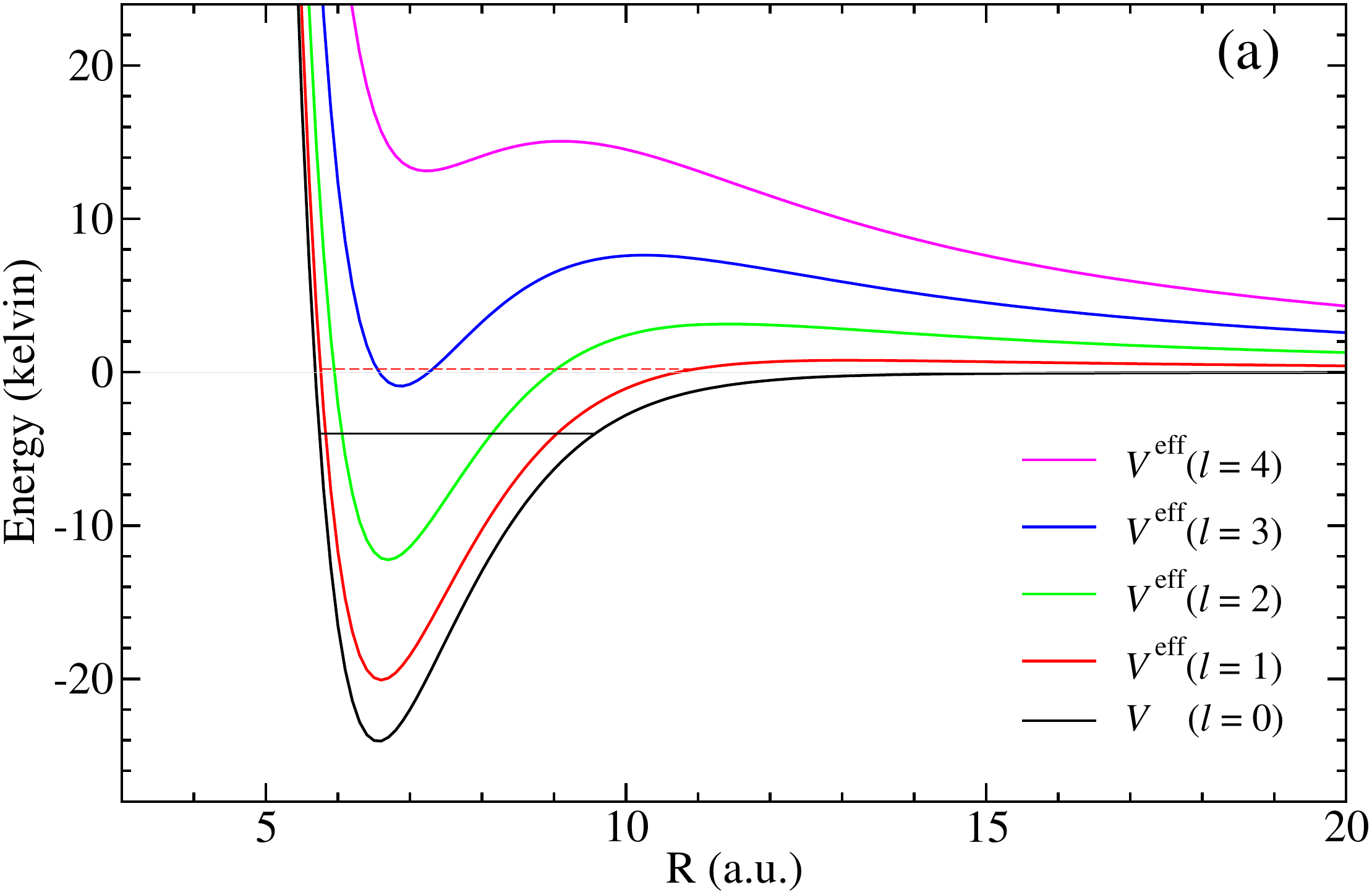}
\hfill
\includegraphics[width=0.47\textwidth]{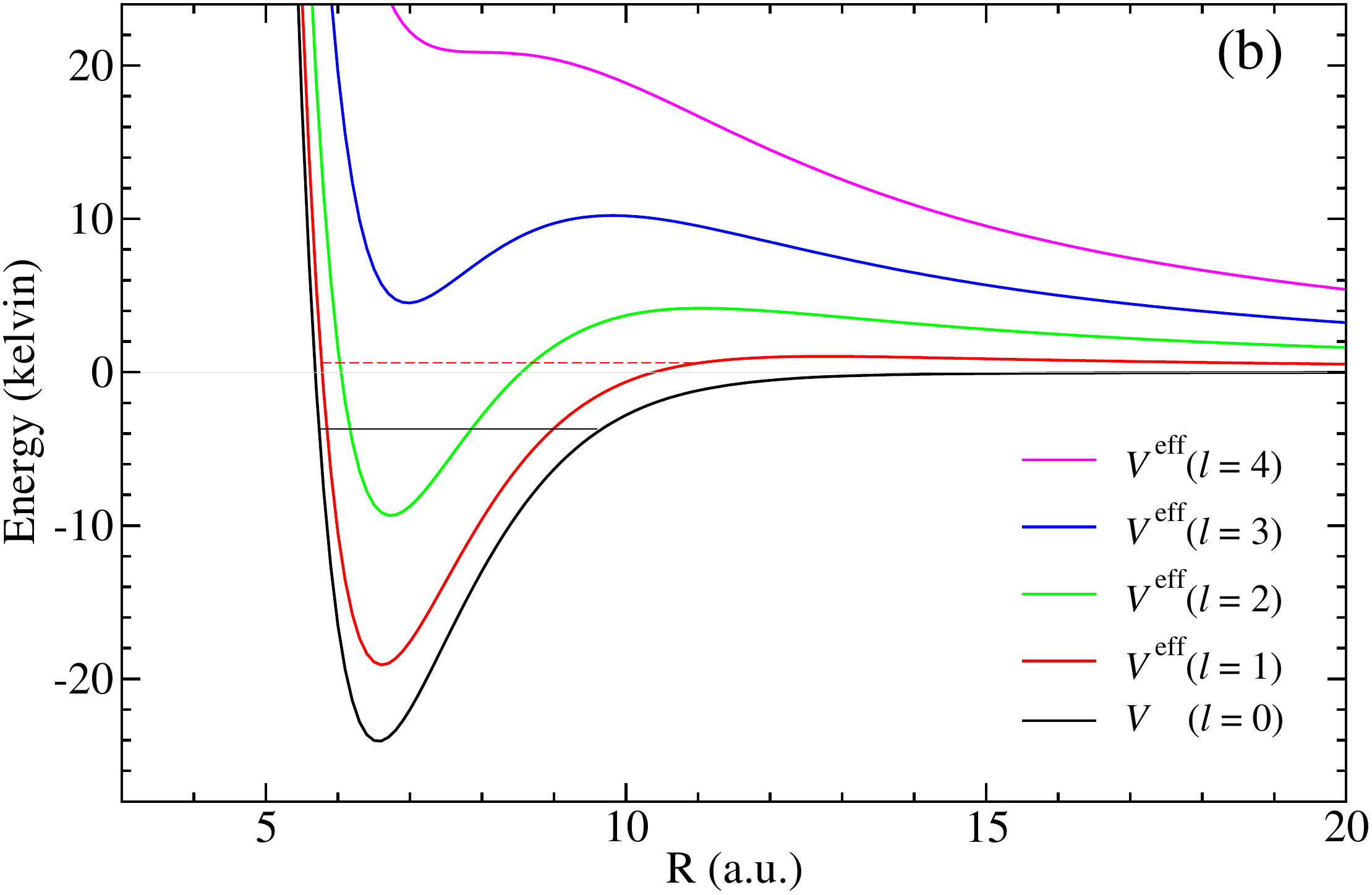}
\caption{
(a) Effective potential, $V^{\rm
    eff}(R)=V(R)+\frac{\ell(\ell+1)}{2\mu R^2}$, for D--H$_2(v=1,j=0)$
  for partial waves $\ell\leq 4$.  The van der Waals well can support
  one bound state (indicated as a horizontal black line) only for
  $\ell=0$.  For $\ell=1$, there is a shape resonance (dashed red
  line) just above the threshold ($E=0$, marked with a grey line). As
  $\ell$ increases, the centrifugal term becomes dominant and prevents
  the formation of bound states, or even the appearance of shape
  resonances.  (b) Same as panel (a), for H---D$_2$.  The
  shape resonance (red, $\ell=1$) is now shifted slightly higher,
  i.e., closer to the top of the centrifugal barrier.  The bound state
  for $\ell=0$ (black) is also shifted higher, slightly closer to the
  threshold.  These differences are mainly due to the centrifugal
  term, $\ell(\ell+1)/2\mu R^2$, which is stronger for H--D$_2$ simply
  because its reduced mass $(\mu\approx 0.8\;\rm u)$ is smaller than
  that of D--H$_2$ $(\mu\approx 1\;\rm u)$, where ``u'' denotes the
  unified atomic mass unit (or dalton).  The unit of length for the
  horizontal axis is the atomic unit (Bohr radius).
\label{fig:V_eff:ell}
}
\end{figure}

Figure~\ref{fig:V_eff:ell} shows the single-channel effective
potential, which includes the centrifugal term corresponding to
partial waves $\ell\leq4$, for initial rovibrational state
$(v,j)=(1,0)$.  For $\ell=1$, the height of the centrifugal barrier is
less than one kelvin, and thus a possible shape resonance may only
appear in the sub-kelvin regime.  Indeed, a shape resonance for p-wave
does exist, as indicated in figure~\ref{fig:V_eff:ell}, and is
confirmed by the full results in figure~\ref{fig:compare:j=0} for both
reactions.  For $\ell=2$, the height of the centrifugal barrier is
approximately 4~K, but the potential well for D$_2$\,+\,H is too
shallow and the d-wave contribution gives small bumps corresponding to
above-barrier resonances in the energy range 6--8~K\@.  However, for
H$_2$\,+\,D, the effective potential for $\ell=2$ is deep enough (due
to the larger reduced mass, which diminishes the centrifugal term) to
allow shape resonances in the energy range 1--4~K\@.
Figure~\ref{fig:compare:j=0} also shows that the resonant features
move towards lower energy as the initial vibrational quantum number
$v$ increases.  In fact, in the case of H$_2$\,+\,D, the p-wave
resonance becomes quasi-bound for $v=2$ and higher, and the binding
energy of the newly formed van der Waals complex increases slightly
with $v$; thus, the corresponding rounded feature in
figure~\ref{fig:compare:j=0}(a,b) moves gradually from the sub-kelvin
regime towards higher energy.

For higher partial waves, the centrifugal term becomes dominant and
the effective potential in figure~\ref{fig:V_eff:ell} is almost
entirely repulsive and thus it cannot bind a van der Waals complex;
consequently, shape resonances cannot appear at energies above
10~K.  Nevertheless, with increasing energy, the contribution of
higher partial waves will become significant.  Note that for $E>100$~K
(not shown here) the rate coefficients increase exponentially for
lower vibrational initial states, as dictated by the reaction barrier;
thus, a rather flat minimum in the energy range 10--100~K is clearly
visible in figure~\ref{fig:compare:j=0}, between the resonant
enhancement at very low energy and the exponential increase at high
energy.

\begin{figure}[t]
\mbox{}\hfill\includegraphics[width=.84\textwidth]{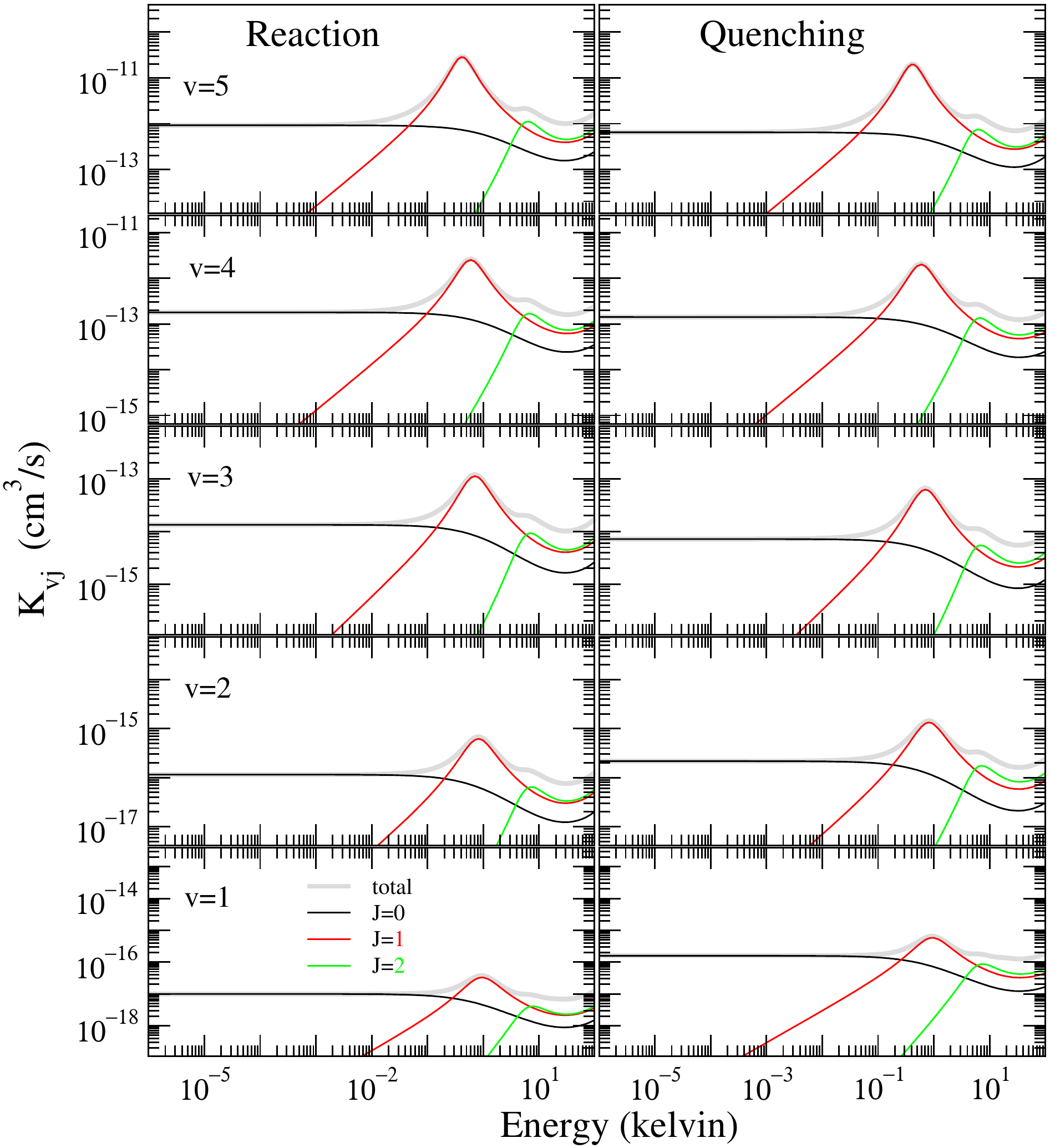}
\caption{
Energy dependent rate coefficients ${\sf K}_{vj}(E)$ for
  H\,+\,D$_2(v,j)$, for initial rovibrational states $v\leq 5$ (as
  indicated) and $j=0$.  The panels on the left correspond to
  reaction, while those on the right to quenching.  Each panel shows
  the total rate (thick gray line) and the individual contributions
  for different values of the total angular momentum $J=0,\;1,\;2$
  (black, red, and green lines).
\label{fig:bigJ:j=0}
}
\end{figure}

Although in this work we only focus on the prominent shape resonances
found at low energy, we remark that the overall magnitude of the rate
coefficients reveals an anomaly; namely, the results for
D$_2(v=1)$\,+\,H are surprisingly large.  For $v\geq 2$, the reaction
rates for D$_2$\,+\,H are lower than the rates for H$_2$\,+\,D, as
expected due to the lower internal vibrational energy of D$_2(v)$
compared to H$_2(v)$.  However, for $v=1$, the background values of
the reaction rates are nearly the same for both reactions.  Moreover,
the rate coefficient for the vibrational quenching of D$_2(v=1)$ is
larger than that of H$_2(v=1)$, if we ignore the shape resonance
(which only affects a limited energy range).
Figure~\ref{fig:compare:j=0} also shows that the quenching rate for
D$_2(v=1)$ is nearly as large as that of D$_2(v=2)$.  We remark that a
simple explanation in terms of Feshbach resonances due to closed
channels, as discussed for example in \cite{FH2-Science} for H$_2$+F
scattering, is ruled out, as there are no nearby channels (open or
closed) to affect the results significantly.  Thus, the $v=1$ anomaly
for D$_2$\,+\,H remains an open question which deserves further study.

Next, we discuss briefly the individual contributions for total
angular momentum $J$\@.  In the case of ortho-D$_2(j$\,=\,$0)$, the
orbital angular momentum in the entrance channel is restricted to
$\ell = J$, and the corresponding sum over $\ell$ in
equation~(\ref{eq:sigma}) reduces to a single term, yielding
\begin{equation}
\label{eq:sigma:j=0}
\sigma_{n'\leftarrow n}(E) = \frac{\pi}{k^{2}_n}
	\sum_{J=0}^{\infty} \left( 2J+1 \right)
    \sum_{\ell'= |J-j'|}^{J+j'}
     \left| S^{J}_{n'\ell', \, nJ}(E)\right|^{2}.
\end{equation} 
For the low energies explored here, only the partial waves $\ell=J\leq
2$ give significant contributions, which are shown in
figure~\ref{fig:bigJ:j=0} for D$_2(v,0)$\,+\,H\@.  At the very lowest
energies, the rate coefficient is dominated by $s$-wave ($\ell=0$).
However, for energies above 100~mK, the s-wave contribution decreases
towards a minimum located at $E\approx 30$~K, as seen in
figure~\ref{fig:bigJ:j=0}.  Thus, higher partial waves become dominant
even in the kelvin regime;  in fact, the resonantly enhanced p-wave
contribution is already dominant in the sub-kelvin regime.

\subsection{Reaction and quenching rates for j\,=\,1}
\label{sec:j=1}

\begin{figure}[t]
\mbox{}\hfill\includegraphics[width=0.84\textwidth]{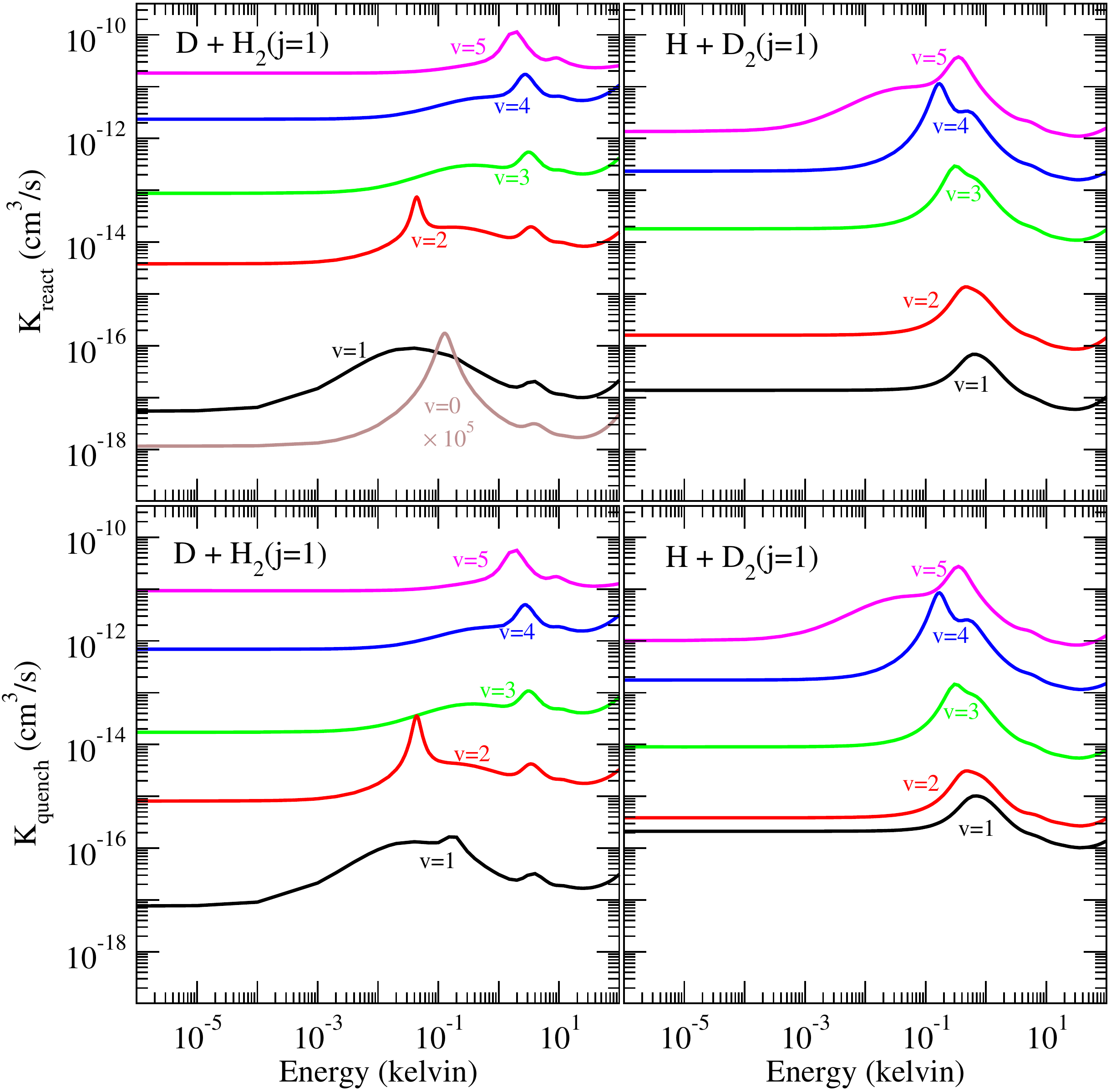}
\caption{Same as figure~\ref{fig:compare:j=0} for initial
  rotational state $j=1$.}
\label{fig:compare:j=1}
\end{figure}

Figure~\ref{fig:compare:j=1} shows the rate coefficients for initial
rotational state $j=1$. We first remark that the resonant features for
$j=1$ are significantly different than the resonances presented in
figure~\ref{fig:compare:j=0} for $j=0$.  Such a strong dependence on
the initial rotational state of the dimer stems mostly from the fact
that different partial waves $(\ell)$ in the entrance channel are
coupled when both $j\ne0$ and $J\ne0$.  Hence, if a shape resonance
in a certain partial wave $\ell$ appears for a particular value $J$ of
the total angular momentum, then it may also manifest for all other
possible values: $J=|\ell-j|,\cdots,\ell+j$.  We recall that the
coupled channel problems for different values of $J$ are separate;
thus, if a certain shape resonance appears repeatedly (for different
values of $J$), each occurrence can be regarded as a different
resonance associated with a particular coupled channel system of
equations for a given $J$.  Moreover, when a shape resonance is near
the entrance channel threshold, it is very sensitive to the details of
the coupled channel problem; e.g., small changes in the coupling
matrix may cause the resonance to shift significantly, either away from
the threshold, or closer to it---even becoming quasi-bound.

\begin{figure}[t]
\mbox{}\hfill\includegraphics[width=0.84\textwidth]{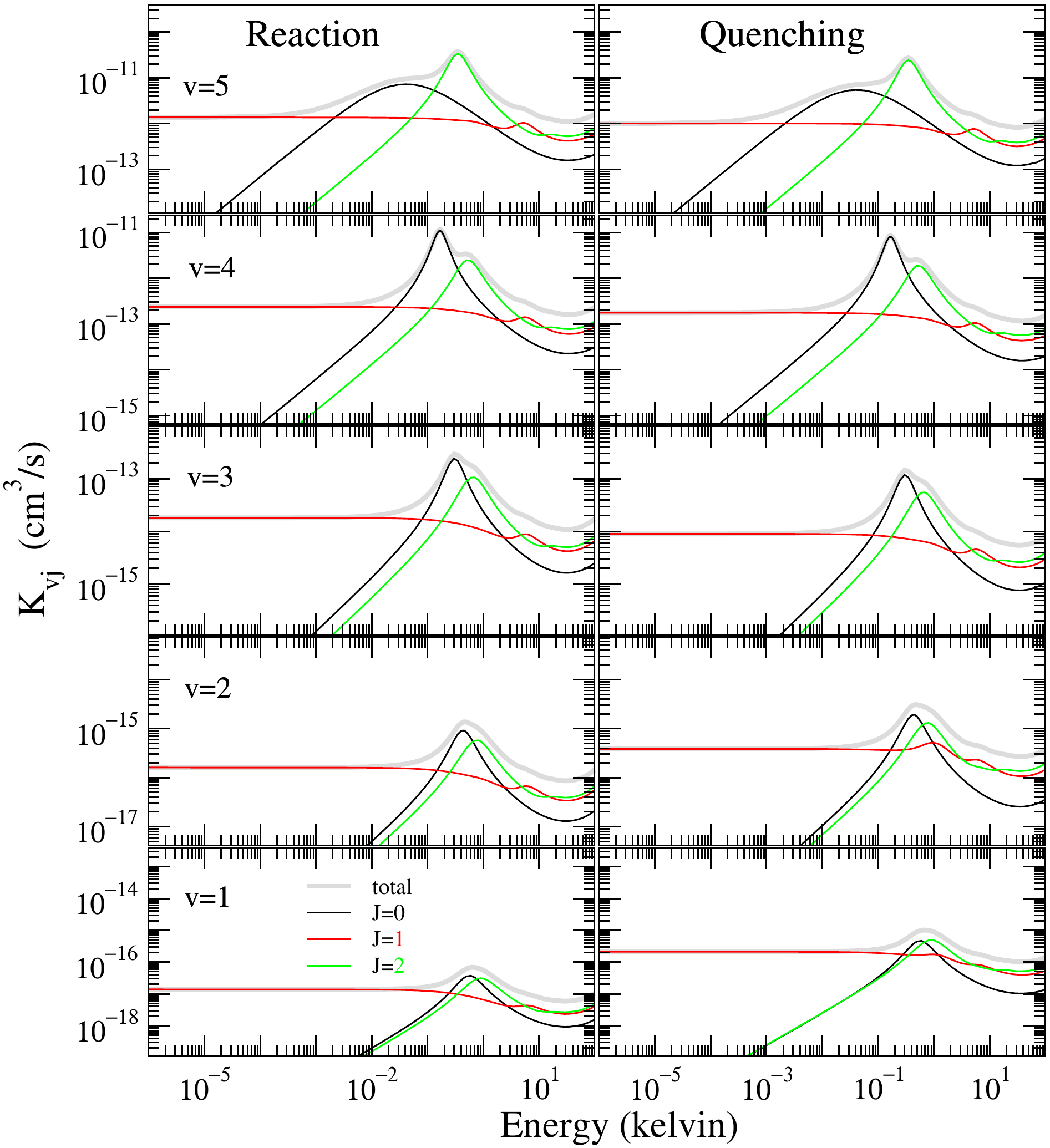}
\caption{Same as figure~\ref{fig:bigJ:j=0}, for
  para-D$_2(v,j)$ with initial rotational state $j=1$ and $v\leq 5$.}
\label{fig:bigJ:j=1}
\end{figure}

To better understand the complicated resonant features of the total
rate coefficients for $j=1$, we show the individual partial $(J)$
contributions in figure~\ref{fig:bigJ:j=1}.  First, note that for
$J=0$, the relative angular momentum $\ell$ in the entrance channel is
restricted to $\ell=j=1$, corresponding to the black curves in
figure~\ref{fig:bigJ:j=1}, which are similar to the results shown
previously in figure~\ref{fig:bigJ:j=0} for $\ell=J=1$.  In both cases
($j=0$ in figure~\ref{fig:bigJ:j=0}, and $j=1$ in
figure~\ref{fig:bigJ:j=1}) there is a gradual shift of the p-wave
shape resonance towards lower energy as $v$ increases.  However, for
$j=1$, the overall attractive effect of the interaction is slightly
stronger than for $j=0$, most notably for $v=5$ with the van der
Waals complex for $j=1$ (with $J=0$ and $\ell=1$) becoming quasi-bound
just below the threshold and producing the rounded resonant feature
shown in figure~\ref{fig:bigJ:j=1} (uppermost panel).  Next, for $J=1$,
the partial waves $\ell=0,1,2$ in the entrance channel are coupled
together, and their net contribution is shown by the red curves in
figure~\ref{fig:bigJ:j=1}. Thus, the $J=1$ contribution includes
s-wave ($\ell=0$, dominant at vanishing energies), p-wave, and d-wave
(responsible for the small bump near $E\approx 6$~K).  For $J=2$, we
have $\ell=1,2,3$, whose contribution is shown by the green curves in
figure~\ref{fig:bigJ:j=1}.  The p-wave shape resonances for $J=0$ and
$J=2$ partly overlap, and thus the sum of all $J$-terms yields a
complicated profile for the total rate coefficients.

\begin{figure}[t]
\mbox{}\hfill\includegraphics[width=0.84\textwidth]{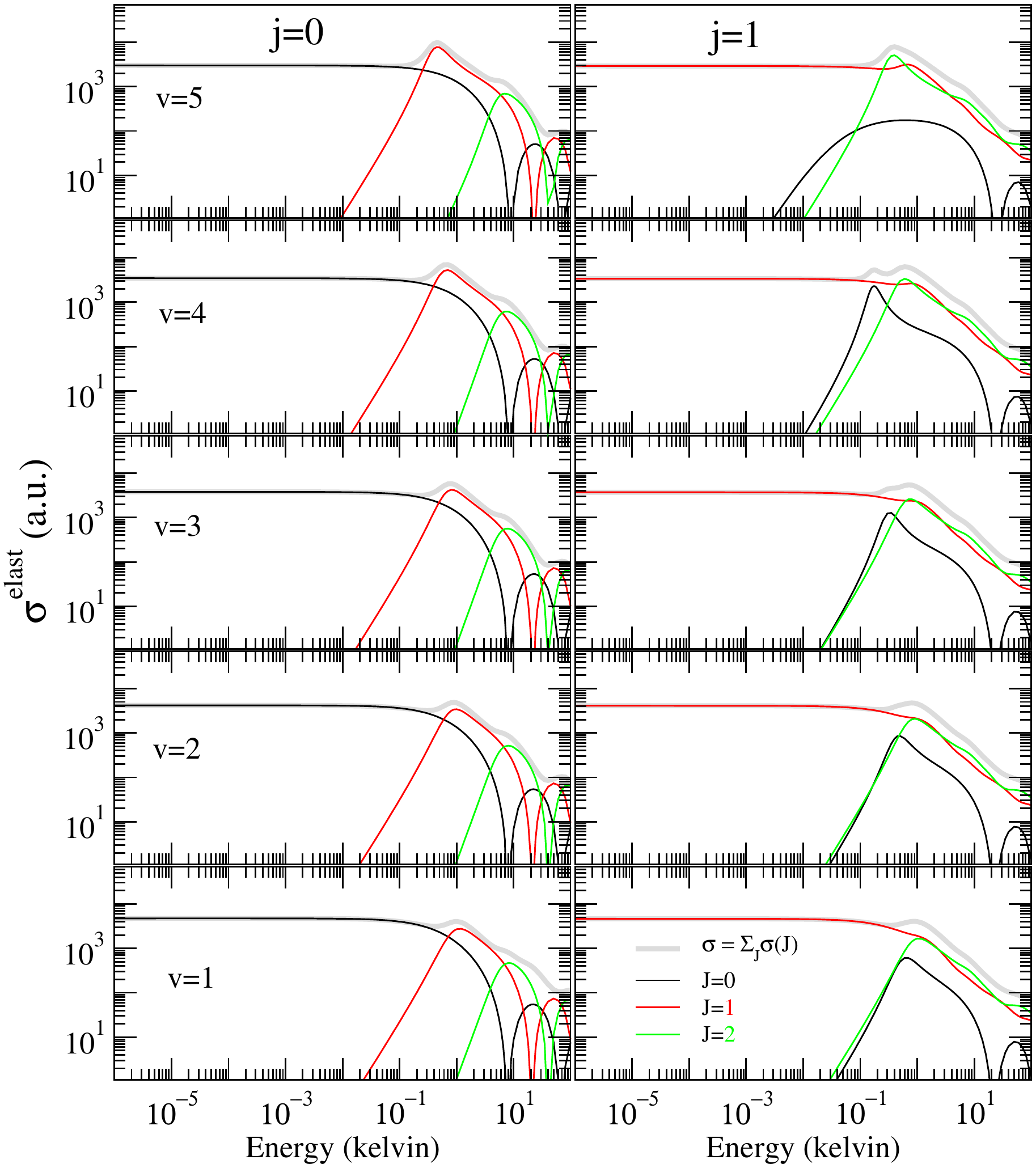}
\caption{Elastic cross sections
  for H\,+\,D$_2(v,j)$ for initial rovibrational states $(v,j)$ with
  $1\leq v\leq5$ and $j=0,\;1$.
   The panels on the left correspond to $j=0$, while those
  on the right side to $j=1$.  Each panel shows the total elastic
  cross section, as well as the contributions of individual
  $J=0,\;1,\;2$.  The atomic unit (Bohr radius squared) is used for
  the cross section.
}
\label{fig:sigma-j-elast}
\end{figure}

\subsection{Elastic cross sections}
\label{sec:elastic}

For the sake of completeness, we describe very briefly the results for
the elastic cross section, $\sigma_{n\leftarrow n}(E)$, which is
denoted as $\sigma^{\rm elast}_{n}(E)$;  its  expression is
obtained by setting $n'=n$ in equation~(\ref{eq:sigma}),
\begin{equation}
\label{eq:sigma-elastic}
 \sigma^{\rm elast}_{n}(E) = \frac{\pi}{k^{2}_n}
	\sum_{J=0}^{\infty} \left(\frac{2J+1}{2j+1}\right)
	\sum_{\ell=|J-j|}^{J+j} \;\; \sum_{\ell'=|J-j|}^{J+j}
     \left| \delta_{\ell'\ell} - S^{J}_{n\ell',\; n\ell}(E)\right|^{2}.
\end{equation}
In general, different partial waves $\ell$ are coupled, and the
elastic cross section comprises the double sum
$\sum_\ell\sum_{\ell'}$, as well as $\sum_J$.  A much simpler
expression for $\sigma^{\rm elast}_{n}(E)$ only exists for initial
rovibrational states $n=(v,j)$ with $j=0$, when we have $\ell=J$ and
$\ell'=J$, and equation~(\ref{eq:sigma-elastic}) reads
\begin{equation}
\label{eq:sigma-elastic-j=0}
    \sigma^{\rm elast}_{v,\,j=0}(E) = \frac{\pi}{k^{2}_{vj}}
	\sum_{\ell=0}^{\infty} \left( 2\ell+1\right)
     \left| 1 - S^{J=\ell}_{vj\ell,\; vj\ell}(E)\right|^{2}.
\end{equation}

The results for $j=0$ and $j=1$ are shown in
figure~\ref{fig:sigma-j-elast} for $1\leq v\leq 5$.  Although the
elastic cross section shows resonant features in the kelvin and
sub-kelvin regimes, they appear less pronounced due to the large value
of the s-wave contribution in the Wigner regime $(E\rightarrow0)$.
The zero energy limit of $\sigma^{\rm elast}_{n}(E)$ decreases from
approximately 5000 to 3000 a.u., while its maximal resonant value near
$E\approx 1$~K increases from 5000 to 10$^4$ a.u., as $v$ increases
from $v=1$ to $v=5$.  Table~\ref{tab1} summarizes the real and
imaginary parts of the scattering length.

\begin{table}[ht]
\caption{\label{tab1}Real and imaginary contributions to the
  scattering length $a_{v,j}=\alpha_{v,j}-i\beta_{v,j}$ extracted from
  the quantum results for the collision of H with ortho-D$_2(j=0)$ and
  para-D$_2(j=1)$.}
\begin{indented}
\lineup
\item[]\begin{tabular}{*{5}{l}}
\br
      &    \centre{2}{H + D$_2(v,j=0)$}    &    \centre{2}{H + D$_2(v,j=1)$} \\
\ns
      &     \crule{2}          &     \crule{2}        \\
\ns
 $v$  &  $\alpha$ (a.u.)  &  $\beta$ (a.u.)  &  $\alpha$ (a.u.)  &  $\beta$
(a.u.)\\
\mr
  1   &   19.4     &  $4.3\times 10^{-6}$    &   19.3   & $3.2\times 10^{-6}$\\
  2   &   18.2     &  $1.0\times 10^{-5}$    &   18.1   & $6.3\times 10^{-6}$\\
  3   &   17.3     &  $5.2\times 10^{-4}$    &   17.2   & $3.3\times 10^{-4}$\\
  4   &   16.6     &  $7.6\times 10^{-3}$    &   16.3   & $6.1\times 10^{-3}$\\
  5   &   15.4     &  $4.5\times 10^{-2}$    &   15.2   & $3.0\times 10^{-2}$\\
\br
\end{tabular}
\end{indented}
\end{table}

\section{Conclusion}
\label{sec:conclusion}

We presented results for the scattering of H on D$_2$ for initial
vibrational states $v\leq5$, and we found prominent shape resonances
at very low energy.  We remark that these resonant structures are
sensitive to the details of the potential energy surface and they
could serve as stringent tests if experimental data would become
available.  We have demonstrated that the initial rotational state of
D$_2$ affects the resonant features significantly;  indeed, the
lineshapes of the resonances for ortho-D$_2(j=0)$ and
para-D$_2(j=1)$ are very different.  Moreover, the initial vibrational
state also has a strong effect on the resonant features.  Finally,
comparing the results presented here for H\,+\,D$_2$ with the results
obtained previously for D\,+\,H$_2$, we report a very strong isotopic
effect, as the resonant features for the two reactions differ drastically.


\section*{Acknowledgments}
This work was partially supported by the MURI US Army Research Office
Grant No. W911NF-14-1-0378 (IS) and by the US Army Research Office,
Chemistry Division, Grant No. W911NF-13-1-0213 (RC).


\section*{References}

\bibliographystyle{my-unsrt}
\bibliography{ref}

\begin{thebibliography}{20}

\bibitem{cote-1997}
R.~C\^ot\'e and A.~Dalgarno, \emph{Mechanism for the production of
  vibrationally excited ultracold molecules of {$^7$Li$_2$}}, Chem. Phys. Lett.
  \textbf{279}, 50 (1997).

\bibitem{jmp-1999}
R.~C\^ot\'e and A.~Dalgarno, \emph{Mechanism for the production of $^6${Li}$_2$
  and $^7${Li}$_2$ ultracold molecules}, J. Mol. Spectr. \textbf{195}, 236
  (1999).

\bibitem{pillet-1998}
A.~Fioretti, D.~Comparat, A.~Crubellier, O.~Dulieu, F.~Masnou-Seeuws, and
  P.~Pillet, \emph{Formation of cold {Cs}$_2$ molecules through
  photoassociation}, Phys. Rev. Lett. \textbf{80}, 4402 (1998).

\bibitem{knize-1998}
T.~Takekoshi, B.~M. Patterson, and R.~J. Knize, \emph{Observation of optically
  trapped cold cesium molecules}, {Phys. Rev. Lett.} \textbf{81}, 5105 (1998).

\bibitem{stwalley:canjchem:04}
W.~C. Stwalley, \emph{Collisions and reactions of ultracold molecules},
  Can.~J.~Chem. \textbf{82}, 709 (2004).

\bibitem{weck+bala:irpc:06}
P.~F. Weck and N.~Balakrishnan, \emph{Importance of long-range interactions in
  chemical reactions at cold and ultracold temperatures}, Int. Rev. Phys. Chem.
  \textbf{25}, 283 (2006).

\bibitem{jeremy:irpc:06}
P.~Sold\'an and J.~M. Hutson, \emph{Molecule formation in ultracold atomic
  gases}, Int. Rev. Phys. Chem. \textbf{25}, 497 (2006).

\bibitem{jeremy:irpc:07}
P.~Sold\'an and J.~M. Hutson, \emph{Molecular collisions in ultracold atomic
  gases}, Int. Rev. Phys. Chem. \textbf{26}, 1 (2007).

\bibitem{krems:pccp:08}
R.~V. Krems, \emph{Cold controlled chemistry}, Phys. Chem. Chem. Phys.
  \textbf{10}, 4079 (2008).

\bibitem{gospel:09}
R.~Krems, W.~Stwalley, and B.~Friedrich, \emph{Cold molecules: theory,
  experiment, applications}, CRC Press (2009).

\bibitem{ian.smith:08}
I.~Smith, \emph{Low temperatures and cold molecules}, Imperial College Press
  (2008).

\bibitem{w+z:09}
M.~Weidem\"uller and C.~Zimmermann, \emph{Cold atoms and molecules: a
  testground for fundamental many particle physics}, Physics textbook,
  Wiley-VCH (2009).

\bibitem{our:NJP-2015:H2D}
I.~Simbotin and R.~C\^ot\'e, \emph{Effect of nuclear spin symmetry in cold and
  ultracold reactions: {D + para/ortho-H$_2$}}, New J. Phys. \textbf{17},
  065003 (2015).

\bibitem{astro}
C.~D. Gay, P.~C. Stancil, S.~Lepp, and A.~Dalgarno, \emph{The highly deuterated
  chemistry of the early universe}, Astrophys. J. \textbf{737}, 44 (2011).

\bibitem{astro-cloud}
D.~Galli and F.~Palla, \emph{Deuterium chemistry in the primordial gas},
  Planet. Space Sci. \textbf{50}, 1197  (2002), {Special} issue on Deuterium in
  the Universe.

\bibitem{Narevicius-2012-Science}
A.~B. {Henson}, S.~{Gersten}, Y.~{Shagam}, J.~{Narevicius}, and
  E.~{Narevicius}, \emph{Observation of resonances in {Penning} ionization
  reactions at sub-kelvin temperatures in merged beams}, Science \textbf{338},
  234 (2012).

\bibitem{Narevicius-2014-NatChem}
E.~{Lavert-Ofir}, Y.~{Shagam}, A.~B. {Henson}, S.~{Gersten}, J.~{K{\l}os},
  P.~S. {{\.Z}uchowski}, J.~{Narevicius}, and E.~{Narevicius},
  \emph{Observation of the isotope effect in sub-kelvin reactions}, Nature
  Chem. \textbf{6}, 332 (2014).

\bibitem{Narevicius-2015-NatChem}
Y.~{Shagam}, A.~{Klein}, W.~{Skomorowski}, R.~{Yun}, V.~{Averbukh}, C.~P.
  {Koch}, and E.~{Narevicius}, \emph{Molecular hydrogen interacts more strongly
  when rotationally excited at low temperatures leading to faster reactions},
  Nature Chem. \textbf{7}, 921 (2015).

\bibitem{Narevicius-2016-NatPhys}
A.~Klein, Y.~Shagam, W.~Skomorowski, P.~S. {\.Z}uchowski, M.~Pawlak, L.~M.~C.
  Janssen, N.~Moiseyev, S.~Y.~T. {van de Meerakker}, A.~{van der Avoird}, C.~P.
  Koch, and N.~E., \emph{Directly probing anisotropy in atom-molecule
  collisions through quantum scattering resonances}, Nature Phys.  (2016),
  doi:10.1038/nphys3904.

\bibitem{abc:cpc:2k}
D.~Skouteris, J.~F. Castillo, and D.~E. Manolopoulos, \emph{{ABC:} a quantum
  reactive scattering program}, Comp. Phys. Comm. \textbf{133}, 128 (2000).

\bibitem{PCCP-H2+D}
I.~Simbotin, S.~Ghosal, and R.~C\^ot\'e, \emph{A case study in ultracold
  reactive scattering: {D + H}$_2$}, Phys. Chem. Chem. Phys. \textbf{13}, 19148
  (2011).

\bibitem{mielke}
S.~L. Mielke, K.~A. Peterson, D.~W. Schwenke, B.~C. Garrett, D.~G. Truhlar,
  J.~V. Michael, M.-C. Su, and J.~W. Sutherland, \emph{{H + H$_{2}$} thermal
  reaction: a convergence of theory and experiment}, Phys. Rev. Lett.
  \textbf{91}, 063201 (2003).

\bibitem{bala-2016}
B.~K. Kendrick, J.~Hazra, and N.~Balakrishnan, \emph{Geometric phase effects in
  the ultracold {D + HD $\rightarrow$ D + HD} and {D + HD $\leftrightarrow$ H +
  D$_2$} reactions}, New J. Phys. \textbf{18}, 123020 (2016).

\bibitem{bkmp2:jcp96}
A.~I. Boothroyd, W.~J. Keogh, P.~G. Martin, and M.~R. Peterson, \emph{A refined
  {H$_3$} potential energy surface}, J. Chem. Phys. \textbf{104}, 7139 (1996).

\bibitem{PRL-rydberg-dressing}
J.~Wang, J.~N. Byrd, I.~Simbotin, and R.~C\^ot\'e, \emph{Tuning ultracold
  chemical reactions via {Rydberg}-dressed interactions}, Phys. Rev. Lett.
  \textbf{113}, 025302 (2014).

\bibitem{our:jost-NTR:ChemPhys-2015}
I.~Simbotin and R.~C\^ot\'e, \emph{Jost function description of near threshold
  resonances for coupled-channel scattering}, Chem. Phys. \textbf{462}, 79
  (2015).

\bibitem{NTR}
I.~Simbotin, S.~Ghosal, and R.~C\^ot\'e, \emph{Threshold resonance effects in
  reactive processes}, Phys. Rev. A \textbf{89}, 040701 (2014).

\bibitem{FH2-Science}
M.~{Qiu}, Z.~{Ren}, L.~{Che}, D.~{Dai}, S.~A. {Harich}, X.~{Wang}, X.~{Yang},
  C.~{Xu}, D.~{Xie}, M.~{Gustafsson}, R.~T. {Skodje}, Z.~{Sun}, and D.~H.
  {Zhang}, \emph{Observation of {Feshbach} resonances in the {F +
  H$_{2}\rightarrow$ HF + H} reaction}, Science \textbf{311}, 1440 (2006).

\end{thebibliography}

\end{document}